\documentclass[twocolumn,showpacs,%
   nofootinbib,aps,superscriptaddress,%
   eqsecnum,prd,notitlepage,showkeys,10pt]{revtex4-1}

\usepackage{amssymb}
\usepackage{amsmath}
\usepackage{graphicx}
\usepackage{dcolumn}
\usepackage{hyperref}
\usepackage{mathtools, cases, bbm, physics,siunitx}

\usepackage[table,xcdraw]{xcolor}
\usepackage{xcolor}

\newcommand{\mi}{\mathrm{i}}

\newcommand{\Ho}{\mathcal{H}}

\newcommand{\chia}{\ket{\chi_{\alpha}}}
\newcommand{\deff}{\vcentcolon=}

\hypersetup{
    colorlinks,
    linkcolor={black!50!black},
    citecolor={black!50!black},
    urlcolor={black!80!black}
}

\begin{document}

\title{Maximizing Time Delays for Toy Models of Quantum Scattering}
\author{Erin Crawley}
\affiliation{University of British Columbia}
\affiliation{Queen's University}

\begin{abstract}

We model a particle entering a complicated system from free space using an infinite chain of simple harmonic oscillators coupled to a finite, $n$-site cluster. For a particle wavepacket with small wavenumber, an expression for the time delay in terms of the coupling strengths of the cluster is found. When the coupling strengths are varied, a minimum and maximum time delay can be found for $n=1$. When $n=2$, we can obtain seemingly arbitrarily large time delays. In both cases, the time delays share similarities with the time delays for the scattering from an analogous quantum well target. We conclude that this could mean that large time delays are caused by interference within the wavefunction in the target's region of space. 
\end{abstract}

\maketitle

\section{Introduction}

In this paper, we consider quantum systems for which energy and matter can become trapped within a localized system for a period of time before eventually escaping. An interesting theoretical example of this type is the problem of Hawking radiation \cite{hawking}. Consider a black hole in otherwise free space; when quantum effects are taken into account, it was theorized by Hawking that the black hole will emit particles as thermal radiation. This causes a decrease in the black hole's mass over time, leading to an eventual disappearance of the black hole. It is clear that in this situation, energy and matter are initially trapped within the (localized) black hole, eventually escaping after some finite amount of time. 

When considering quantum systems which trap energy and matter, a few questions naturally arise. Firstly, we want to obtain an expression for the time delay, the amount of time for which the matter is trapped within the localized system. Moreover, we wish to examine how these types of systems can be set up, and whether we can vary parameters of the system in order to maximize the time delay. 

While these broad questions are somewhat difficult to answer for a general quantum system, they can be answered for some toy models. Moreover, the consideration of these toy models could help to further understanding of general quantum systems which trap energy and matter in a localized space. In this paper, we consider a localized, $n$-site cluster connected to a discretized version of 1-dimensional free space. Using this model, we will find an expression for the time delay in the case of general $n$, and attempt to maximize the time delay for the $n=1$ and $n=2$ cases. 
 
\section{The Model}

To model this situation, we need an open system which describes free particles, which we then couple to a finite, more complex quantum system. 

\begin{figure}[h]
\includegraphics[width=8cm]{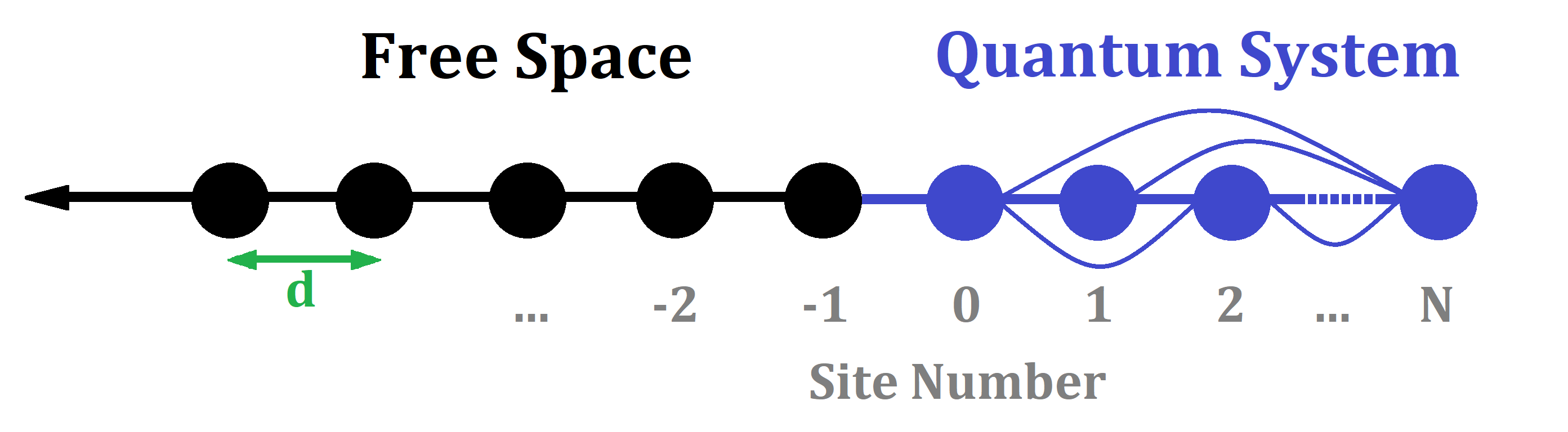}
\caption{\label{fig_model} The proposed model. The spacing between adjacent sites is $d$. }
\end{figure}

To model the open system, we consider an infinite chain of simple harmonic oscillators (SHOs) coupled by nearest-neighbour interactions. We couple this system to a more complex, finite system of SHOs, which have possible couplings to all other finite system SHOs, as shown in Fig. \ref{fig_model}. The Hamiltonian for the total system, $\Ho$, is taken to be:

\begin{equation}
\Ho=\Ho_0 + \Ho_1, 
\end{equation}
where
\begin{gather}
\Ho_0 = A \sum_{n\leq0} a_n^\dagger a_n + B \sum_{n\leq-1} (a_n^\dagger a_{n-1} + a_{n-1}^\dagger a_{n}),\\
\text{and }
\Ho_1 = \sum_{i,j=0}^N g_{ij}a_i^\dagger a_j,\\
\text{with } g_{ij} = g^*_{ji}.
\end{gather}

\subsection{Eigenstates of $\Ho_0$}

Let $\ket{0}$ be the 0-particle eigenstate of the number operator, $\mathcal{N} =  \sum_j a_j^\dagger a_j$. It can be shown that $[\mathcal{N}, \Ho_0]=0$, so there exists a basis of $\Ho_0$ eigenstates with definite particle number.

The single-particle eigenstates of $\Ho_0$ are:
\begin{equation}
\ket{\chi_\alpha} = \sum_{n\leq -1} (e^{i \alpha n}-e^{-i \alpha n}) a_n^\dagger \ket{0} \label{eq_chia}
\end{equation}

The corresponding energy eigenvalues are:
\begin{equation}
E_\alpha = A+2B\cos(\alpha), \alpha \in \mathbb{R}
\end{equation}
From the correspondence between the momentum operator and translation \cite{marksnotes}, we can show that $\alpha = kd$, where $k$ is the wavenumber and $d$ is the spacing between SHO sites. In the following, we will consider results for small $\alpha$. 

We note that the eigenstates given by \eqref{eq_chia} are discretized versions of free particle states. So, we can choose
\begin{equation} \label{eq_1dshoinfin_smallklimit_AB}
\begin{aligned}
A&=\frac{\hbar^2}{m d^2},\\ 
B&=-\frac{\hbar^2}{2 m d^2},
\end{aligned}
\end{equation}
such that for small $\alpha$, the energy of a free particle with wavenumber $k=\tfrac{\alpha}{d}$ is $E_\alpha=\tfrac{p^2}{2m}=\tfrac{\hbar^2 \alpha^2}{2md^2}$.

\subsection{Eigenstates of $\Ho$}
%\note{mention wavepackets built from these modes}
Far from the added complex sites (i.e. for free particle sites with site number $n<<0$), the effect of these complex sites should be diminished. So, the energy eigenvalues of $\Ho$ should be $E_\alpha$, with similar eigenstates to $\chia$. Suppose the eigenstates of $\Ho$ are of the form
\begin{equation}
\ket{\Psi_\alpha} = \sum_{n\leq -1} (e^{\mi \alpha n} +e^{-\mi \alpha n}e^{\mi \varphi(\alpha)})a_n^\dagger \ket{0} + \sum_{n = 0}^N \beta_n(\alpha) a_n^\dagger \ket{0}.    \label{eq_psialpha}
\end{equation}

Letting $G$ be the matrix obtained from the coupling strengths, we can define
\begin{displaymath}
P :=\left(E_\alpha \operatorname{I}_{N\times N} - G \right)=
\begin{bmatrix}
    E_\alpha-g_{00} &  \hdots & -g_{0N} \\
    \vdots           &  \ddots &         \vdots   \\
    -g_{N0} & \hdots & E_\alpha-g_{NN} \\
\end{bmatrix},
\end{displaymath}
\begin{displaymath}
 \text{and, if $P$ is invertible, } 
P^{-1} :=
\begin{bmatrix}
    \mu_{00}(\alpha) &  \hdots & \mu_{0N}(\alpha) \\
    \vdots           &  \ddots &         \vdots   \\
    \mu_{N0}(\alpha) & \hdots & \mu_{NN}(\alpha) \\
\end{bmatrix}.
\end{displaymath}
From the definition of $g_{ij}$, $P$ is a Hermitian matrix. It can also be shown that if $P$ is invertible, $P^{-1}$ is Hermitian. 

Since $\Ho$ has energy eigenvalues $E_\alpha$, restricting that $\Ho \ket{\Psi_\alpha} = E_\alpha \ket{\Psi_\alpha}$ yields
\begin{equation}
e^{\mi \varphi(\alpha)} = \frac{-\left(1-B\mu_{00}(\alpha)e^{-\mi \alpha}\right)}{1-B\mu_{00}(\alpha)e^{\mi \alpha}}. \label{eq_eiphi}
\end{equation}

% Then, \eqref{eq_psialpha} becomes:

% \begin{equation}
% \begin{split}
% \ket{\Psi_\alpha} = \sum_{n\leq -1} (e^{\mi \alpha n} +e^{-\mi \alpha n}\frac{-\left(1-B\mu_{00}(\alpha)e^{-\mi \alpha}\right)}{1-B\mu_{00}(\alpha)e^{\mi \alpha}})a_n^\dagger \ket{0} \\
% + \sum_{n = 0}^N \beta_n(\alpha) a_n^\dagger \ket{0}    \label{eq_psialphanew}
% \end{split}
% \end{equation}

\subsection{Particle Wavepackets}
With this construction, we can now consider how to model a particle coming in from infinity and entering the complex system. In this model, we consider a particle as a Gaussian wave packet of width $\Delta$ centred at $ \alpha_0 $, given by

\begin{equation}
\ket{\psi_{\alpha_0,\Delta}(t)} = \int_{-\infty}^{\infty} \dd{\alpha} e^{\frac{(\alpha-\alpha_0)^2}{\Delta^2}} e^{\frac{-\mi E_\alpha t}{\hbar} } \ket{\Psi_\alpha}, \label{eq_wavepacket}
\end{equation}
where $\ket{\Psi_\alpha}$ is as in \eqref{eq_psialpha}.

We will consider narrow wavepackets, where $\Delta<<1$. In this case, \eqref{eq_wavepacket} is well approximated by:
\begin{equation}
\ket{\psi_{\alpha_0,\Delta}(t)} = \int_{-\epsilon}^{\epsilon} \dd{\alpha} e^{\frac{(\alpha-\alpha_0)^2}{\Delta^2}} e^{\frac{-\mi E_\alpha t}{\hbar} } \ket{\Psi_\alpha} \label{eq_wavepacketnew}
\end{equation}
for some small $\epsilon$. Thus, our small $\alpha$ condition for the particle is still retained for the wavepacket \eqref{eq_wavepacket}. Moreover, at early times, this wavepacket is localized in space. 

\section{The Time Delay}
Since $\Delta$ is small, we can consider the Taylor series approximation of $\varphi(\alpha)$ about $\alpha_0$. This expansion yields:
\begin{displaymath}
\begin{aligned}
\varphi(\alpha)&\approx \varphi_0+\varphi_1(\alpha-\alpha_0),\\ 
\text{where }\varphi_0&\deff \varphi(\alpha_0)\text{ and } \varphi_1\deff \dv{\varphi(\alpha)}{\alpha}\bigg|_{\alpha_0}
\end{aligned}
\end{displaymath}

We can show that the centre of the packet, $X$, moves according to 
\begin{displaymath}
\begin{cases}
X=\frac{\hbar\alpha_0}{md} t, \text{ for } t<<0\\
X=-\frac{\hbar \alpha_0}{md} \left( t -\frac{md\varphi_1}{\hbar\alpha_0}\right), \text{ for } t>>0
\end{cases}
\end{displaymath}

This can be interpreted as the particle moving towards the complex system with group velocity $v_g=\frac{\hbar \alpha_0}{md}$, then much later in time moving in the opposite direction, away from the system, with the same speed. The particle experiences a delay of $\frac{md\varphi_1}{\hbar\alpha_0}$ between entering and leaving the complex system. So, the time, $\tau$, that the particle can be said to stay within the complex system is
\begin{equation}
\tau \deff \tfrac{m d^2 \varphi'(\alpha_0)}{\hbar\alpha_0}. \label{eq_tau}
\end{equation}

Further discussion on how to find time delays for quantum system scattering can be found in \cite{timedelay}. Defining dimensionless quantities $\tau_* \deff \tfrac{\hbar}{2 m d^2} \tau$ and $\tilde{\mu}_{ij} := B \mu_{ij}$, and combining \eqref{eq_eiphi} and \eqref{eq_tau} yields:
\small
\begin{equation}
\tau_* =  \frac{2 \alpha_0 \left[\sum_{n=0}^N |\tilde{\mu}_{0n}(\alpha_0)|^2\right] \sin(\alpha_0)+ \tilde{\mu}_{00}(\alpha_0) \cos(\alpha_0) - \tilde{\mu}^2_{00}(\alpha_0)}{\alpha_0\left(1 -2\tilde{\mu}_{00}(\alpha_0)\cos(\alpha_0)+\tilde{\mu}^2_{00}(\alpha_0)\right)}  \label{eq_taustar}
\end{equation}
\normalsize
\eqref{eq_taustar} is the (scaled) time that the particle spends within the complex system. 

\section{Results}
We can now consider the time delay for some specific examples, as well as attempt find systems which maximize $\tau$. Since our expression for the time delay depends on the entries of an $(N+1)\times (N+1)$ matrix, the computational difficulty increases considerably as $N$ increases. So, we will consider the cases of one or two added complex sites. 

\subsection{One Complex Site}

For one complex site, the time delay is given by:

\begin{equation}
\tau_*=\frac{1}{\alpha_0} \left[\frac{1 -(2+\frac{g_{00}}{B})\cos(\alpha_0)}{1-2(2+\frac{g_{00}}{B})\cos(\alpha_0)+(2+\frac{g_{00}}{B})^2}\right] \label{eq_td_Nis0_tauexpression}
\end{equation}

It can be shown that the graph of $\tau_*$ vs. $\frac{g_{00}}{B}$ will attain a unique maximum and minimum. \footnote{In fact, letting $C\deff \cos(\alpha_0)$, the maximum: $$\tau_*^{max}= \frac{1}{2\alpha_0} \frac{ C } {\sqrt{\frac{1}{C^2}-1} - (\tfrac{1}{C} -C) }$$ is attained at $(\frac{g_{00}}{B})^{max} \deff \tfrac{1}{C}-2 - \sqrt{\tfrac{1}{C^2}-1}$, and the minimum: $$\tau_*^{min}= \frac{1}{2\alpha_0} \frac{ -C } {\sqrt{\tfrac{1}{C^2}-1} + (\tfrac{1}{C} -C) }$$ is attained at $(\frac{g_{00}}{B})^{min} \deff \tfrac{1}{C}-2 + \sqrt{\tfrac{1}{C^2}-1}$. }

We can plot $\tfrac{\tau_*}{t_*} $ vs. $\frac{g_{00}}{B}$. The time delay, $\tau_*$, is scaled by $t_* \deff \tfrac{\hbar}{2 m d^2}v_gd=\tfrac{\hbar}{2 m d^2}\tfrac{md^2}{\hbar \alpha_0}=\tfrac{1}{2\alpha_0}$, the (scaled) time required for the particle moving at $v_g$ to move one site distance, $d$. The dimensionless quantity we plot is thus $\tfrac{\tau_*}{t_*} = 2\alpha_0\tau_*$

\begin{figure}[h]
\includegraphics[width=8cm]{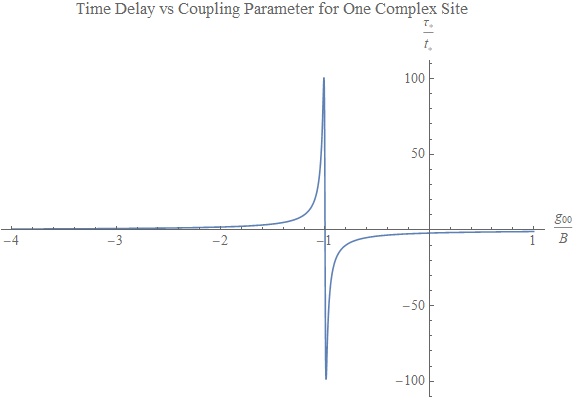}
\caption{\label{fig_onesite} The time delay versus $\tfrac{g_{00}}{B}$ for $\alpha_0 = 0.01$ for one complex site.}
\end{figure}

A plot of $\tfrac{\tau_*}{t_*} $ vs. $\frac{g_{00}}{B}$ where $\alpha_0 = 0.01$ is shown in Fig. \ref{fig_onesite}. We note that for other small values of $\alpha_0$, the qualitative behaviour of the plot appears to remain the same.

\subsection{Two Complex Sites}
For two added sites, we consider the parameters $g_{00}$, $g_{11}$, $g_{01}$, and $g_{10} = g^*_{01}$. However, by examining the expression for $\tau_*$, we see that without loss of generality, we can let $g_{01}$ be real, as $\tau_*$ is a function of $g_{00}$, $g_{11}$, and $|g_{01}|$. 

\begin{figure}[h]
\includegraphics[width=8cm]{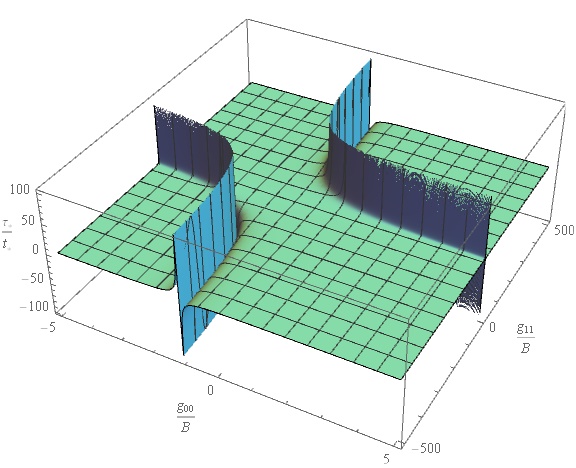}
\caption{\label{fig_twosite} The time delay vs $\tfrac{g_{00}}{B}$ and $\tfrac{g_{11}}{B}$ for $\alpha_0 = 0.01$ and $\tfrac{g_{01}}{B} = 10$.}
\end{figure}

It was determined qualitatively that changes in $g_{01}$ had less of an effect on $\tau_*$. Fig. \ref{fig_twosite} shows the time delay versus $\tfrac{g_{00}}{B}$ and $\tfrac{g_{11}}{B}$ for fixed $\tfrac{g_{01}}{B}=10$. 

\begin{figure}[h!]
\includegraphics[width=6.5cm]{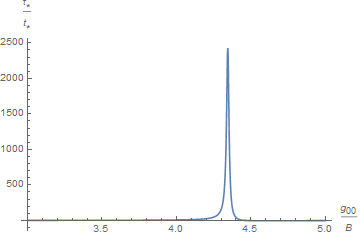}
\caption[cap1]{\label{fig_max1} The time delay vs $\tfrac{g_{00}}{B}$ for $\alpha_0 = 0.01$ and fixed $\tfrac{g_{01}}{B} = 0.2172918414991763, \tfrac{g_{11}}{B} = 0.008733615838984361$.}
\end{figure}

\begin{figure}[h!]
\includegraphics[width=6.5cm]{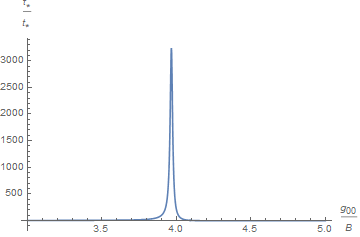}
\caption{\label{fig_max2} The time delay vs $\tfrac{g_{00}}{B}$ for $\alpha_0 = 0.01$ and fixed $\tfrac{g_{01}}{B} = 0.17468678350764077, \tfrac{g_{11}}{B} = 0.006044726011816858$.}
\end{figure}

\begin{figure}[h!]
\includegraphics[width=6.5cm]{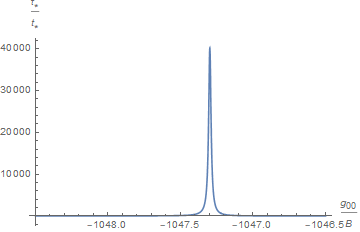}
\caption{\label{fig_max3} The time delay vs $\tfrac{g_{00}}{B}$ for $\alpha_0 = 0.01$ and fixed $\tfrac{g_{01}}{B} = 10.431652777523526, \tfrac{g_{11}}{B} = -0.10410419882719132$.}
\end{figure}

\begin{figure}[h!]
\includegraphics[width=6.5cm]{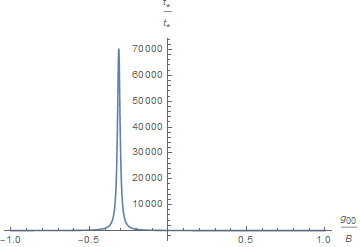}
\caption{\label{fig_max4} The time delay vs $\tfrac{g_{00}}{B}$ for $\alpha_0 = 0.01$ and fixed $\tfrac{g_{01}}{B} = 0.005175467209083724, \tfrac{g_{11}}{B} = -0.00006096640188476812$.}
\end{figure}

We note the similar qualitative features of the two site case to that of the one site case. However, in contrast to the one site case, for two added sites, it appears that we can obtain arbitrarily large time delays for specifically tuned $(g_{00}, g_{01}, g_{11})$ couplings. That is, there does not seem to be a specific $(g_{00}, g_{01}, g_{11})$ point where $\tau_*$ is maximized. This can be seen from Figs. \ref{fig_max1} -- \ref{fig_max4}, which show some peak time delays for specifically chosen coupling parameters.


% \begin{table}[]
% \begin{tabular}{|l|l|l|l|}
% \hline
% $\frac{g_{00}}{B}$  & $\frac{g_{11}}{B}$ & $|\frac{g_{01}}{B}|=|\frac{g_{10}}{B}|$ & Maximum Time Delay, $\tau_*$ \\ \hline
% $-11.012$           & -0.1               & 1                                       & $2.131 \times 10^4$          \\ \hline
% $-1.002\times 10^3$ & -0.1               & 10                                      & $2.004\times 10^6$           \\ \hline
% $-1.002\times 10^5$ & -0.1               & 100                                     & $2.004\times 10^8$           \\ \hline
% $-1.002\times 10^7$ & -0.1               & 1000                                    & $2.004\times 10^{10}$        \\ \hline
% \end{tabular}
% \end{table}

\newpage
\section{Discussion}
\subsection{Connection Between the One Site Case and a 1D Quantum Well} \label{sec_1sitewell}

\begin{figure}[h]
\includegraphics[width=4.5cm]{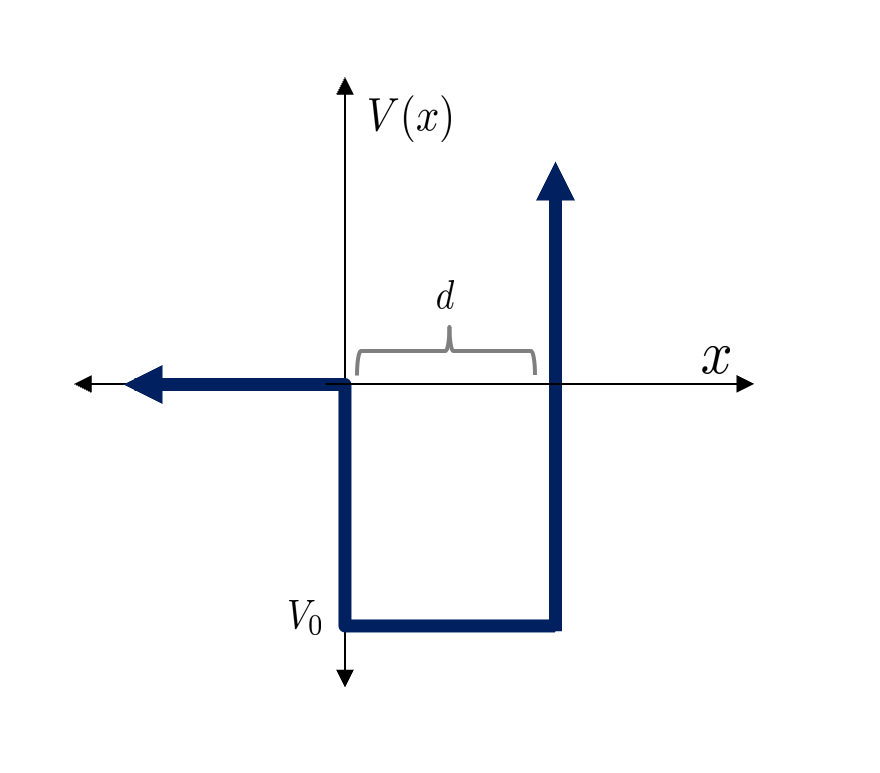}
\caption{\label{fig_well1} The potential for a 1D quantum well of width $d$.}
\end{figure}

Consider a particle beam $A_0e^{ikx}$ incident on a potential barrier given by:
\begin{displaymath}
V(x) = 
\begin{cases}
0 \text{ , if } x<0 \\
V_0 \text{ , if } 0<x<L \\
\infty \text{ , if } x>L \\
\end{cases}
\end{displaymath}
The graph of this potential is shown in Fig. \ref{fig_well1}. 

\begin{figure}[h]
\includegraphics[width=8cm]{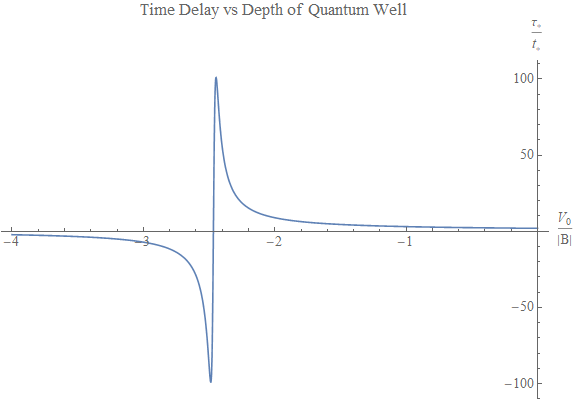}
\caption{\label{fig_wellgraph1} The scaled time delay, $\frac{\tau_*}{t_*}$ versus scaled well depth, $\frac{V_0}{|B|}$}
\end{figure}

\begin{figure}[h]
\includegraphics[width=8cm]{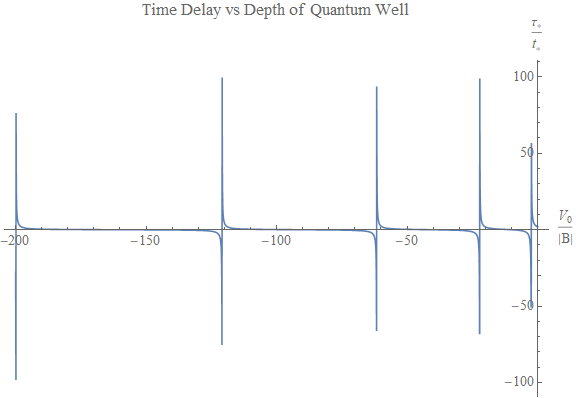}
\caption{\label{fig_wellgraphB} The scaled time delay, $\frac{\tau_*}{t_*}$ versus scaled well depth, $\frac{V_0}{|B|}$ for a longer $\frac{V_0}{|B|}$ scale than Fig. \ref{fig_wellgraph1}.}
\end{figure}

By solving this elementary problem, we can obtain an expression for $R$, the complex amplitude of the reflected wave, which is of the form $R = A_0 e^{\mi \phi(k)}=A_0 e^{\mi \varphi(\alpha)}$. Then, an expression for the time delay can be found, as $\tau_*=\frac{ \varphi'(\alpha_0)}{2 \alpha_0}$. Some plots of the time delay versus well depth are shown in Fig. \ref{fig_wellgraph1} and Fig. \ref{fig_wellgraphB}. 

We note that there are qualitative similarities in the graphs of $\tau_*$ vs. $g_{00}$ and that of $\tau_*$ vs. $V_0$ when $\frac{g_{00}}{B}$ and $\frac{V_{0}}{|B|}$ are on similar scales. So, we take $V_0$ to be a quantum scattering analogue of $g_{00}$. 

\subsection{Connection to Thin Film Problems}

\begin{figure}[ht]
\includegraphics[width=6.5cm]{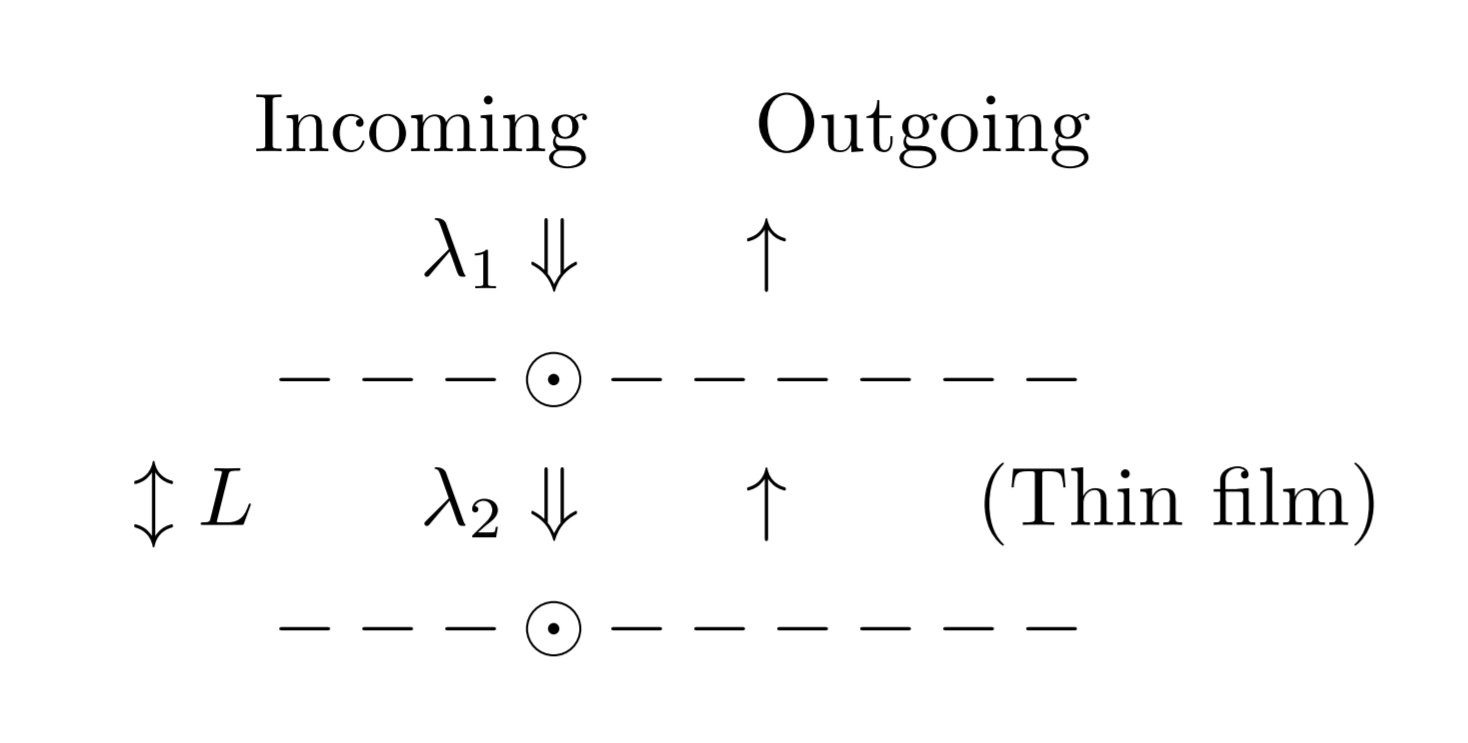}
\caption{\label{fig_thinfilm} An illustration of a thin film problem.}
\end{figure}

Now, consider a thin film, where a wave is partially transmitted and partially reflected at each boundary, as in Fig. \ref{fig_thinfilm}. 

In a 1D quantum well, $k_1<k_2$, so we have constructive interference if $\frac{V_0}{|B|} = k^2L^2 - \frac{(m + \tfrac{1}{2})^2}{4}$, and destructive interference if $\frac{V_0}{|B|} = k^2L^2 - \frac{m^2}{4}$, where $m \in \mathbb{Z}$. 

We note that the time delays in Fig. \ref{fig_wellgraphB} occur at approximately quadratically-spaced intervals. Thus, the maxima in the quantum well time delay case could be caused by interference of the wavefunction within the well. While the quantum well is not a perfect analogue to the one site case, the two agree quite well when $\frac{g_{00}}{B}$ and $\frac{V_{0}}{|B|}$ are on similar scales. So, constructive and destructive interference could also be an explanation for the time delay maxima and minima seen in the one site case of the toy model. 

\subsection{Connection to a Quantum Well for Two Sites}

Now, we examine if the quantum-well analogue will extend to the case of two complex sites.

Consider a particle beam $A_0e^{ikx}$ incident on the potential given by:
\begin{displaymath}
V(x) = 
\begin{cases}
0 \text{ , if } x<0 \\
V_0 \text{ , if } 0<x<a \\
V_1 \text{ , if } a<x<2a \\
\infty \text{ , if } x>2a \\
\end{cases}
\end{displaymath}

We can carry out the same process of Section \ref{sec_1sitewell}, solving for $R$ as $R = A_0 e^{\mi \varphi(\alpha)}$, then using $\varphi(\alpha)$ to find an expression for the time delay. 

\begin{figure}[h!]
\includegraphics[width=8cm]{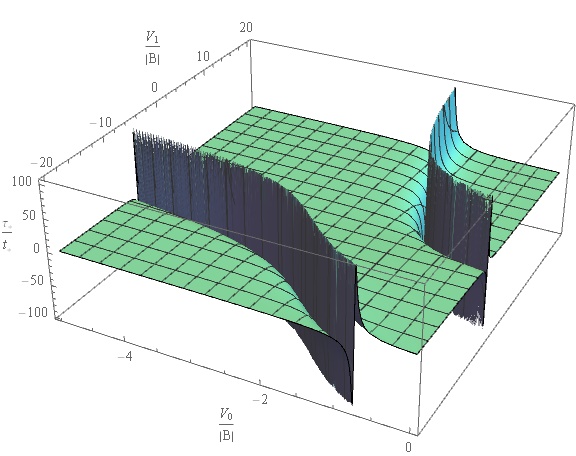}
\caption{\label{fig_wellgraph2} The scaled time delay, $\frac{\tau_*}{t_*}$ versus scaled well depths, $\frac{V_0}{|B|}$ and $\frac{V_1}{|B|}$.}
\end{figure}

\begin{figure}[h!]
\includegraphics[width=8cm]{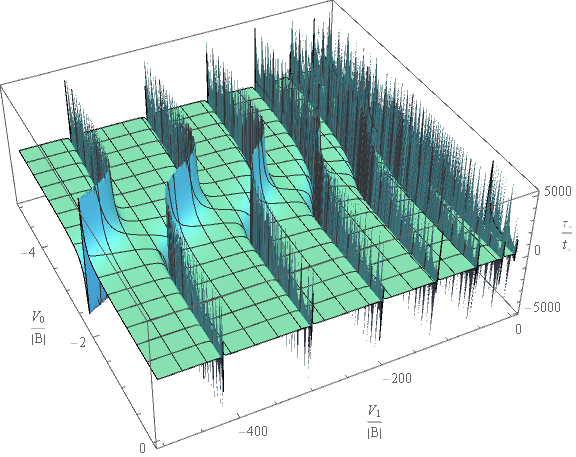}
\caption{\label{fig_wellgraph3} The scaled time delay, $\frac{\tau_*}{t_*}$ versus scaled well depths, $\frac{V_0}{|B|}$ and $\frac{V_1}{|B|}$ for a larger time scale than in Fig. \ref{fig_wellgraph2}.}
\end{figure}

Some plots of the time delay as a function of well depths are shown in Fig. \ref{fig_wellgraph2} and Fig. \ref{fig_wellgraph3}. 

For the double quantum well, the plot of time delay versus well depths shows qualitative similarities to the plot of the time delay versus $g_{00}$ and $g_{11}$ in the two site case. Moreover, for the double quantum well, we can also obtain seemingly arbitrarily large time delays for specifically chosen $(V_0, V_1)$ points. So, it seems as though a quantum well is also a good analogue for the two site case. Then, we propose that similarly to the one site case, the large maxima in the two site case could be due to interference in the particle's wavefunction within the complex sites. \\
\newline

\subsection{Maximizing $\tau_*$ for Fixed Coupling Strengths}
Now, we consider a general number of added sites. Instead of varying the strengths of the couplings, we consider fixing the quantum site coupling strengths so that 
\begin{displaymath}
{g_{ij} := 
\begin{cases}
A, \text{ for } i=j\\
B \text{ or } 0, \text{ for } i\neq j.
\end{cases}}
\end{displaymath}

That is, the complex sites are coupled to one another with the same strengths as in the free particle sites, with the option of complex sites being coupled to sites other than just the nearest-neighbours. 

\begin{figure}[h]
\includegraphics[width=8cm]{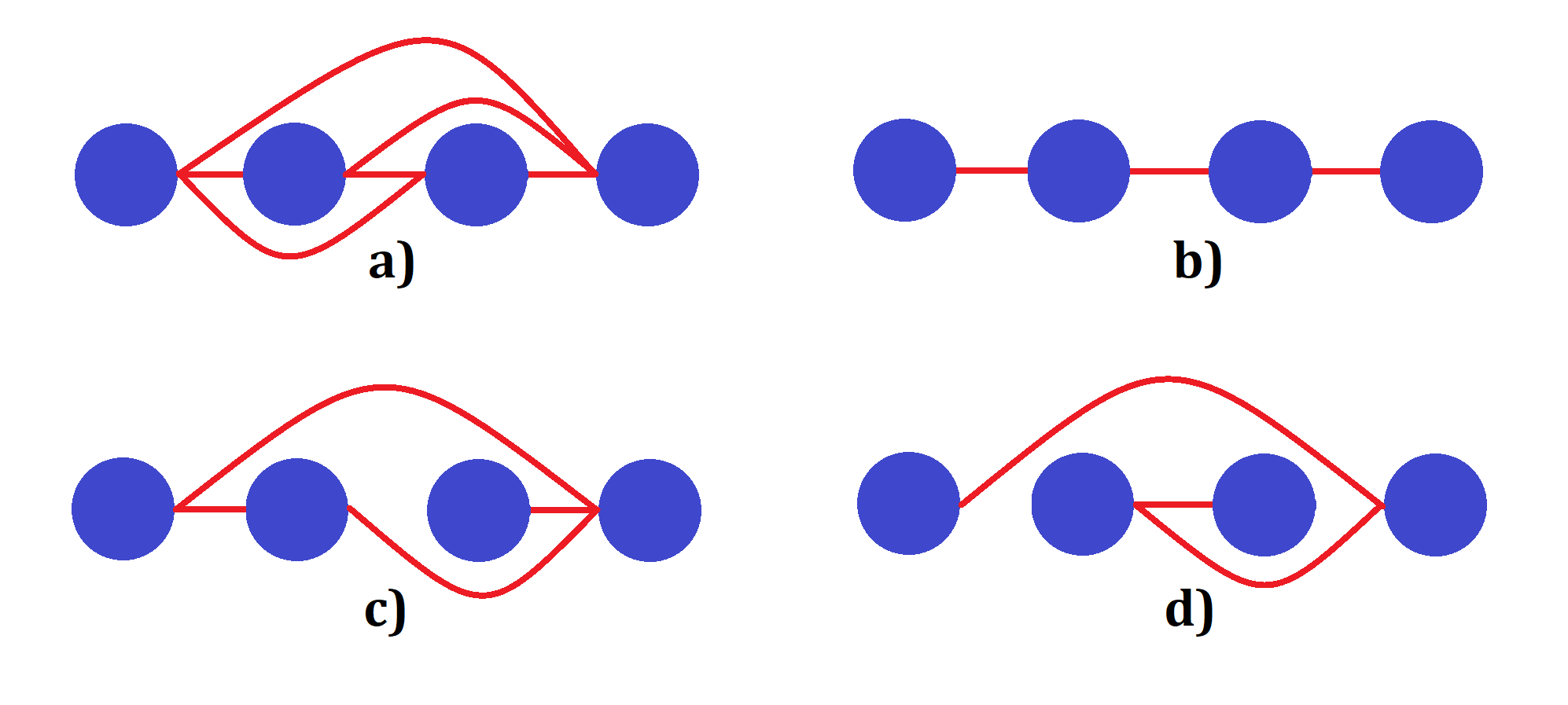}
\caption{\label{fig_fixed_couples} Some example coupling schemes for four complex sites.}
\end{figure}

Some possible coupling schemes for the case of four complex sites is shown in Fig. \ref{fig_fixed_couples}.  

Examined up to $N=6$, the largest time delay occurs when oscillators are simply “chained” end to end, as in b) or d) of Fig. \ref{fig_fixed_couples}. So, for fixed coupling strengths, we cannot obtain a larger time delay than the one obtained by only connecting sites to two neighbours.

\begin{acknowledgments}
This work would not have been possible without the guidance of my supervisor, Mark Van Raamsdonk. Additionally, thanks go to the University of British Columbia Department of Physics and Astronomy and the Natural Sciences and Engineering Research Council of Canada for providing funding for this research. 
\end{acknowledgments}


\begin{thebibliography}{9}
\bibitem{hawking}
S.W. Hawking, Commun. Math. Phys. (1975) 43: 199. https://doi.org/10.1007/BF02345020

\bibitem{marksnotes} 
M. Van Raamsdonk, \textit{Notes on Quantum Mechanics}. http://www.phas.ubc.ca/~mav/Phys402/Notes.pdf (2018)

\bibitem{timedelay}
M. Sassoli de Bianchi, Open Physics. (2011) Vol 10.2: 282. https://doi.org/10.2478/s11534-011-0105-5.

\end{thebibliography}
\end{document}